\documentclass{article}

\makeatletter
\def\PYGdefault@reset{\let\PYGdefault@it=\relax \let\PYGdefault@bf=\relax%
    \let\PYGdefault@ul=\relax \let\PYGdefault@tc=\relax%
    \let\PYGdefault@bc=\relax \let\PYGdefault@ff=\relax}
\def\PYGdefault@tok#1{\csname PYGdefault@tok@#1\endcsname}
\def\PYGdefault@toks#1+{\ifx\relax#1\empty\else%
    \PYGdefault@tok{#1}\expandafter\PYGdefault@toks\fi}
\def\PYGdefault@do#1{\PYGdefault@bc{\PYGdefault@tc{\PYGdefault@ul{%
    \PYGdefault@it{\PYGdefault@bf{\PYGdefault@ff{#1}}}}}}}
\def\PYGdefault#1#2{\PYGdefault@reset\PYGdefault@toks#1+\relax+\PYGdefault@do{#2}}

\expandafter\def\csname PYGdefault@tok@w\endcsname{\def\PYGdefault@tc##1{\textcolor[rgb]{0.73,0.73,0.73}{##1}}}
\expandafter\def\csname PYGdefault@tok@c\endcsname{\let\PYGdefault@it=\textit\def\PYGdefault@tc##1{\textcolor[rgb]{0.25,0.50,0.50}{##1}}}
\expandafter\def\csname PYGdefault@tok@cp\endcsname{\def\PYGdefault@tc##1{\textcolor[rgb]{0.74,0.48,0.00}{##1}}}
\expandafter\def\csname PYGdefault@tok@k\endcsname{\let\PYGdefault@bf=\textbf\def\PYGdefault@tc##1{\textcolor[rgb]{0.00,0.50,0.00}{##1}}}
\expandafter\def\csname PYGdefault@tok@kp\endcsname{\def\PYGdefault@tc##1{\textcolor[rgb]{0.00,0.50,0.00}{##1}}}
\expandafter\def\csname PYGdefault@tok@kt\endcsname{\def\PYGdefault@tc##1{\textcolor[rgb]{0.69,0.00,0.25}{##1}}}
\expandafter\def\csname PYGdefault@tok@o\endcsname{\def\PYGdefault@tc##1{\textcolor[rgb]{0.40,0.40,0.40}{##1}}}
\expandafter\def\csname PYGdefault@tok@ow\endcsname{\let\PYGdefault@bf=\textbf\def\PYGdefault@tc##1{\textcolor[rgb]{0.67,0.13,1.00}{##1}}}
\expandafter\def\csname PYGdefault@tok@nb\endcsname{\def\PYGdefault@tc##1{\textcolor[rgb]{0.00,0.50,0.00}{##1}}}
\expandafter\def\csname PYGdefault@tok@nf\endcsname{\def\PYGdefault@tc##1{\textcolor[rgb]{0.00,0.00,1.00}{##1}}}
\expandafter\def\csname PYGdefault@tok@nc\endcsname{\let\PYGdefault@bf=\textbf\def\PYGdefault@tc##1{\textcolor[rgb]{0.00,0.00,1.00}{##1}}}
\expandafter\def\csname PYGdefault@tok@nn\endcsname{\let\PYGdefault@bf=\textbf\def\PYGdefault@tc##1{\textcolor[rgb]{0.00,0.00,1.00}{##1}}}
\expandafter\def\csname PYGdefault@tok@ne\endcsname{\let\PYGdefault@bf=\textbf\def\PYGdefault@tc##1{\textcolor[rgb]{0.82,0.25,0.23}{##1}}}
\expandafter\def\csname PYGdefault@tok@nv\endcsname{\def\PYGdefault@tc##1{\textcolor[rgb]{0.10,0.09,0.49}{##1}}}
\expandafter\def\csname PYGdefault@tok@no\endcsname{\def\PYGdefault@tc##1{\textcolor[rgb]{0.53,0.00,0.00}{##1}}}
\expandafter\def\csname PYGdefault@tok@nl\endcsname{\def\PYGdefault@tc##1{\textcolor[rgb]{0.63,0.63,0.00}{##1}}}
\expandafter\def\csname PYGdefault@tok@ni\endcsname{\let\PYGdefault@bf=\textbf\def\PYGdefault@tc##1{\textcolor[rgb]{0.60,0.60,0.60}{##1}}}
\expandafter\def\csname PYGdefault@tok@na\endcsname{\def\PYGdefault@tc##1{\textcolor[rgb]{0.49,0.56,0.16}{##1}}}
\expandafter\def\csname PYGdefault@tok@nt\endcsname{\let\PYGdefault@bf=\textbf\def\PYGdefault@tc##1{\textcolor[rgb]{0.00,0.50,0.00}{##1}}}
\expandafter\def\csname PYGdefault@tok@nd\endcsname{\def\PYGdefault@tc##1{\textcolor[rgb]{0.67,0.13,1.00}{##1}}}
\expandafter\def\csname PYGdefault@tok@s\endcsname{\def\PYGdefault@tc##1{\textcolor[rgb]{0.73,0.13,0.13}{##1}}}
\expandafter\def\csname PYGdefault@tok@sd\endcsname{\let\PYGdefault@it=\textit\def\PYGdefault@tc##1{\textcolor[rgb]{0.73,0.13,0.13}{##1}}}
\expandafter\def\csname PYGdefault@tok@si\endcsname{\let\PYGdefault@bf=\textbf\def\PYGdefault@tc##1{\textcolor[rgb]{0.73,0.40,0.53}{##1}}}
\expandafter\def\csname PYGdefault@tok@se\endcsname{\let\PYGdefault@bf=\textbf\def\PYGdefault@tc##1{\textcolor[rgb]{0.73,0.40,0.13}{##1}}}
\expandafter\def\csname PYGdefault@tok@sr\endcsname{\def\PYGdefault@tc##1{\textcolor[rgb]{0.73,0.40,0.53}{##1}}}
\expandafter\def\csname PYGdefault@tok@ss\endcsname{\def\PYGdefault@tc##1{\textcolor[rgb]{0.10,0.09,0.49}{##1}}}
\expandafter\def\csname PYGdefault@tok@sx\endcsname{\def\PYGdefault@tc##1{\textcolor[rgb]{0.00,0.50,0.00}{##1}}}
\expandafter\def\csname PYGdefault@tok@m\endcsname{\def\PYGdefault@tc##1{\textcolor[rgb]{0.40,0.40,0.40}{##1}}}
\expandafter\def\csname PYGdefault@tok@gh\endcsname{\let\PYGdefault@bf=\textbf\def\PYGdefault@tc##1{\textcolor[rgb]{0.00,0.00,0.50}{##1}}}
\expandafter\def\csname PYGdefault@tok@gu\endcsname{\let\PYGdefault@bf=\textbf\def\PYGdefault@tc##1{\textcolor[rgb]{0.50,0.00,0.50}{##1}}}
\expandafter\def\csname PYGdefault@tok@gd\endcsname{\def\PYGdefault@tc##1{\textcolor[rgb]{0.63,0.00,0.00}{##1}}}
\expandafter\def\csname PYGdefault@tok@gi\endcsname{\def\PYGdefault@tc##1{\textcolor[rgb]{0.00,0.63,0.00}{##1}}}
\expandafter\def\csname PYGdefault@tok@gr\endcsname{\def\PYGdefault@tc##1{\textcolor[rgb]{1.00,0.00,0.00}{##1}}}
\expandafter\def\csname PYGdefault@tok@ge\endcsname{\let\PYGdefault@it=\textit}
\expandafter\def\csname PYGdefault@tok@gs\endcsname{\let\PYGdefault@bf=\textbf}
\expandafter\def\csname PYGdefault@tok@gp\endcsname{\let\PYGdefault@bf=\textbf\def\PYGdefault@tc##1{\textcolor[rgb]{0.00,0.00,0.50}{##1}}}
\expandafter\def\csname PYGdefault@tok@go\endcsname{\def\PYGdefault@tc##1{\textcolor[rgb]{0.53,0.53,0.53}{##1}}}
\expandafter\def\csname PYGdefault@tok@gt\endcsname{\def\PYGdefault@tc##1{\textcolor[rgb]{0.00,0.27,0.87}{##1}}}
\expandafter\def\csname PYGdefault@tok@err\endcsname{\def\PYGdefault@bc##1{\setlength{\fboxsep}{0pt}\fcolorbox[rgb]{1.00,0.00,0.00}{1,1,1}{\strut ##1}}}
\expandafter\def\csname PYGdefault@tok@kc\endcsname{\let\PYGdefault@bf=\textbf\def\PYGdefault@tc##1{\textcolor[rgb]{0.00,0.50,0.00}{##1}}}
\expandafter\def\csname PYGdefault@tok@kd\endcsname{\let\PYGdefault@bf=\textbf\def\PYGdefault@tc##1{\textcolor[rgb]{0.00,0.50,0.00}{##1}}}
\expandafter\def\csname PYGdefault@tok@kn\endcsname{\let\PYGdefault@bf=\textbf\def\PYGdefault@tc##1{\textcolor[rgb]{0.00,0.50,0.00}{##1}}}
\expandafter\def\csname PYGdefault@tok@kr\endcsname{\let\PYGdefault@bf=\textbf\def\PYGdefault@tc##1{\textcolor[rgb]{0.00,0.50,0.00}{##1}}}
\expandafter\def\csname PYGdefault@tok@bp\endcsname{\def\PYGdefault@tc##1{\textcolor[rgb]{0.00,0.50,0.00}{##1}}}
\expandafter\def\csname PYGdefault@tok@fm\endcsname{\def\PYGdefault@tc##1{\textcolor[rgb]{0.00,0.00,1.00}{##1}}}
\expandafter\def\csname PYGdefault@tok@vc\endcsname{\def\PYGdefault@tc##1{\textcolor[rgb]{0.10,0.09,0.49}{##1}}}
\expandafter\def\csname PYGdefault@tok@vg\endcsname{\def\PYGdefault@tc##1{\textcolor[rgb]{0.10,0.09,0.49}{##1}}}
\expandafter\def\csname PYGdefault@tok@vi\endcsname{\def\PYGdefault@tc##1{\textcolor[rgb]{0.10,0.09,0.49}{##1}}}
\expandafter\def\csname PYGdefault@tok@vm\endcsname{\def\PYGdefault@tc##1{\textcolor[rgb]{0.10,0.09,0.49}{##1}}}
\expandafter\def\csname PYGdefault@tok@sa\endcsname{\def\PYGdefault@tc##1{\textcolor[rgb]{0.73,0.13,0.13}{##1}}}
\expandafter\def\csname PYGdefault@tok@sb\endcsname{\def\PYGdefault@tc##1{\textcolor[rgb]{0.73,0.13,0.13}{##1}}}
\expandafter\def\csname PYGdefault@tok@sc\endcsname{\def\PYGdefault@tc##1{\textcolor[rgb]{0.73,0.13,0.13}{##1}}}
\expandafter\def\csname PYGdefault@tok@dl\endcsname{\def\PYGdefault@tc##1{\textcolor[rgb]{0.73,0.13,0.13}{##1}}}
\expandafter\def\csname PYGdefault@tok@s2\endcsname{\def\PYGdefault@tc##1{\textcolor[rgb]{0.73,0.13,0.13}{##1}}}
\expandafter\def\csname PYGdefault@tok@sh\endcsname{\def\PYGdefault@tc##1{\textcolor[rgb]{0.73,0.13,0.13}{##1}}}
\expandafter\def\csname PYGdefault@tok@s1\endcsname{\def\PYGdefault@tc##1{\textcolor[rgb]{0.73,0.13,0.13}{##1}}}
\expandafter\def\csname PYGdefault@tok@mb\endcsname{\def\PYGdefault@tc##1{\textcolor[rgb]{0.40,0.40,0.40}{##1}}}
\expandafter\def\csname PYGdefault@tok@mf\endcsname{\def\PYGdefault@tc##1{\textcolor[rgb]{0.40,0.40,0.40}{##1}}}
\expandafter\def\csname PYGdefault@tok@mh\endcsname{\def\PYGdefault@tc##1{\textcolor[rgb]{0.40,0.40,0.40}{##1}}}
\expandafter\def\csname PYGdefault@tok@mi\endcsname{\def\PYGdefault@tc##1{\textcolor[rgb]{0.40,0.40,0.40}{##1}}}
\expandafter\def\csname PYGdefault@tok@il\endcsname{\def\PYGdefault@tc##1{\textcolor[rgb]{0.40,0.40,0.40}{##1}}}
\expandafter\def\csname PYGdefault@tok@mo\endcsname{\def\PYGdefault@tc##1{\textcolor[rgb]{0.40,0.40,0.40}{##1}}}
\expandafter\def\csname PYGdefault@tok@ch\endcsname{\let\PYGdefault@it=\textit\def\PYGdefault@tc##1{\textcolor[rgb]{0.25,0.50,0.50}{##1}}}
\expandafter\def\csname PYGdefault@tok@cm\endcsname{\let\PYGdefault@it=\textit\def\PYGdefault@tc##1{\textcolor[rgb]{0.25,0.50,0.50}{##1}}}
\expandafter\def\csname PYGdefault@tok@cpf\endcsname{\let\PYGdefault@it=\textit\def\PYGdefault@tc##1{\textcolor[rgb]{0.25,0.50,0.50}{##1}}}
\expandafter\def\csname PYGdefault@tok@c1\endcsname{\let\PYGdefault@it=\textit\def\PYGdefault@tc##1{\textcolor[rgb]{0.25,0.50,0.50}{##1}}}
\expandafter\def\csname PYGdefault@tok@cs\endcsname{\let\PYGdefault@it=\textit\def\PYGdefault@tc##1{\textcolor[rgb]{0.25,0.50,0.50}{##1}}}

\makeatother

\makeatletter
\def\PYG@reset{\let\PYG@it=\relax \let\PYG@bf=\relax%
    \let\PYG@ul=\relax \let\PYG@tc=\relax%
    \let\PYG@bc=\relax \let\PYG@ff=\relax}
\def\PYG@tok#1{\csname PYG@tok@#1\endcsname}
\def\PYG@toks#1+{\ifx\relax#1\empty\else%
    \PYG@tok{#1}\expandafter\PYG@toks\fi}
\def\PYG@do#1{\PYG@bc{\PYG@tc{\PYG@ul{%
    \PYG@it{\PYG@bf{\PYG@ff{#1}}}}}}}
\def\PYG#1#2{\PYG@reset\PYG@toks#1+\relax+\PYG@do{#2}}

\expandafter\def\csname PYG@tok@w\endcsname{\def\PYG@tc##1{\textcolor[rgb]{0.73,0.73,0.73}{##1}}}
\expandafter\def\csname PYG@tok@c\endcsname{\let\PYG@it=\textit\def\PYG@tc##1{\textcolor[rgb]{0.25,0.50,0.50}{##1}}}
\expandafter\def\csname PYG@tok@cp\endcsname{\def\PYG@tc##1{\textcolor[rgb]{0.74,0.48,0.00}{##1}}}
\expandafter\def\csname PYG@tok@k\endcsname{\let\PYG@bf=\textbf\def\PYG@tc##1{\textcolor[rgb]{0.00,0.50,0.00}{##1}}}
\expandafter\def\csname PYG@tok@kp\endcsname{\def\PYG@tc##1{\textcolor[rgb]{0.00,0.50,0.00}{##1}}}
\expandafter\def\csname PYG@tok@kt\endcsname{\def\PYG@tc##1{\textcolor[rgb]{0.69,0.00,0.25}{##1}}}
\expandafter\def\csname PYG@tok@o\endcsname{\def\PYG@tc##1{\textcolor[rgb]{0.40,0.40,0.40}{##1}}}
\expandafter\def\csname PYG@tok@ow\endcsname{\let\PYG@bf=\textbf\def\PYG@tc##1{\textcolor[rgb]{0.67,0.13,1.00}{##1}}}
\expandafter\def\csname PYG@tok@nb\endcsname{\def\PYG@tc##1{\textcolor[rgb]{0.00,0.50,0.00}{##1}}}
\expandafter\def\csname PYG@tok@nf\endcsname{\def\PYG@tc##1{\textcolor[rgb]{0.00,0.00,1.00}{##1}}}
\expandafter\def\csname PYG@tok@nc\endcsname{\let\PYG@bf=\textbf\def\PYG@tc##1{\textcolor[rgb]{0.00,0.00,1.00}{##1}}}
\expandafter\def\csname PYG@tok@nn\endcsname{\let\PYG@bf=\textbf\def\PYG@tc##1{\textcolor[rgb]{0.00,0.00,1.00}{##1}}}
\expandafter\def\csname PYG@tok@ne\endcsname{\let\PYG@bf=\textbf\def\PYG@tc##1{\textcolor[rgb]{0.82,0.25,0.23}{##1}}}
\expandafter\def\csname PYG@tok@nv\endcsname{\def\PYG@tc##1{\textcolor[rgb]{0.10,0.09,0.49}{##1}}}
\expandafter\def\csname PYG@tok@no\endcsname{\def\PYG@tc##1{\textcolor[rgb]{0.53,0.00,0.00}{##1}}}
\expandafter\def\csname PYG@tok@nl\endcsname{\def\PYG@tc##1{\textcolor[rgb]{0.63,0.63,0.00}{##1}}}
\expandafter\def\csname PYG@tok@ni\endcsname{\let\PYG@bf=\textbf\def\PYG@tc##1{\textcolor[rgb]{0.60,0.60,0.60}{##1}}}
\expandafter\def\csname PYG@tok@na\endcsname{\def\PYG@tc##1{\textcolor[rgb]{0.49,0.56,0.16}{##1}}}
\expandafter\def\csname PYG@tok@nt\endcsname{\let\PYG@bf=\textbf\def\PYG@tc##1{\textcolor[rgb]{0.00,0.50,0.00}{##1}}}
\expandafter\def\csname PYG@tok@nd\endcsname{\def\PYG@tc##1{\textcolor[rgb]{0.67,0.13,1.00}{##1}}}
\expandafter\def\csname PYG@tok@s\endcsname{\def\PYG@tc##1{\textcolor[rgb]{0.73,0.13,0.13}{##1}}}
\expandafter\def\csname PYG@tok@sd\endcsname{\let\PYG@it=\textit\def\PYG@tc##1{\textcolor[rgb]{0.73,0.13,0.13}{##1}}}
\expandafter\def\csname PYG@tok@si\endcsname{\let\PYG@bf=\textbf\def\PYG@tc##1{\textcolor[rgb]{0.73,0.40,0.53}{##1}}}
\expandafter\def\csname PYG@tok@se\endcsname{\let\PYG@bf=\textbf\def\PYG@tc##1{\textcolor[rgb]{0.73,0.40,0.13}{##1}}}
\expandafter\def\csname PYG@tok@sr\endcsname{\def\PYG@tc##1{\textcolor[rgb]{0.73,0.40,0.53}{##1}}}
\expandafter\def\csname PYG@tok@ss\endcsname{\def\PYG@tc##1{\textcolor[rgb]{0.10,0.09,0.49}{##1}}}
\expandafter\def\csname PYG@tok@sx\endcsname{\def\PYG@tc##1{\textcolor[rgb]{0.00,0.50,0.00}{##1}}}
\expandafter\def\csname PYG@tok@m\endcsname{\def\PYG@tc##1{\textcolor[rgb]{0.40,0.40,0.40}{##1}}}
\expandafter\def\csname PYG@tok@gh\endcsname{\let\PYG@bf=\textbf\def\PYG@tc##1{\textcolor[rgb]{0.00,0.00,0.50}{##1}}}
\expandafter\def\csname PYG@tok@gu\endcsname{\let\PYG@bf=\textbf\def\PYG@tc##1{\textcolor[rgb]{0.50,0.00,0.50}{##1}}}
\expandafter\def\csname PYG@tok@gd\endcsname{\def\PYG@tc##1{\textcolor[rgb]{0.63,0.00,0.00}{##1}}}
\expandafter\def\csname PYG@tok@gi\endcsname{\def\PYG@tc##1{\textcolor[rgb]{0.00,0.63,0.00}{##1}}}
\expandafter\def\csname PYG@tok@gr\endcsname{\def\PYG@tc##1{\textcolor[rgb]{1.00,0.00,0.00}{##1}}}
\expandafter\def\csname PYG@tok@ge\endcsname{\let\PYG@it=\textit}
\expandafter\def\csname PYG@tok@gs\endcsname{\let\PYG@bf=\textbf}
\expandafter\def\csname PYG@tok@gp\endcsname{\let\PYG@bf=\textbf\def\PYG@tc##1{\textcolor[rgb]{0.00,0.00,0.50}{##1}}}
\expandafter\def\csname PYG@tok@go\endcsname{\def\PYG@tc##1{\textcolor[rgb]{0.53,0.53,0.53}{##1}}}
\expandafter\def\csname PYG@tok@gt\endcsname{\def\PYG@tc##1{\textcolor[rgb]{0.00,0.27,0.87}{##1}}}
\expandafter\def\csname PYG@tok@err\endcsname{\def\PYG@bc##1{\setlength{\fboxsep}{0pt}\fcolorbox[rgb]{1.00,0.00,0.00}{1,1,1}{\strut ##1}}}
\expandafter\def\csname PYG@tok@kc\endcsname{\let\PYG@bf=\textbf\def\PYG@tc##1{\textcolor[rgb]{0.00,0.50,0.00}{##1}}}
\expandafter\def\csname PYG@tok@kd\endcsname{\let\PYG@bf=\textbf\def\PYG@tc##1{\textcolor[rgb]{0.00,0.50,0.00}{##1}}}
\expandafter\def\csname PYG@tok@kn\endcsname{\let\PYG@bf=\textbf\def\PYG@tc##1{\textcolor[rgb]{0.00,0.50,0.00}{##1}}}
\expandafter\def\csname PYG@tok@kr\endcsname{\let\PYG@bf=\textbf\def\PYG@tc##1{\textcolor[rgb]{0.00,0.50,0.00}{##1}}}
\expandafter\def\csname PYG@tok@bp\endcsname{\def\PYG@tc##1{\textcolor[rgb]{0.00,0.50,0.00}{##1}}}
\expandafter\def\csname PYG@tok@fm\endcsname{\def\PYG@tc##1{\textcolor[rgb]{0.00,0.00,1.00}{##1}}}
\expandafter\def\csname PYG@tok@vc\endcsname{\def\PYG@tc##1{\textcolor[rgb]{0.10,0.09,0.49}{##1}}}
\expandafter\def\csname PYG@tok@vg\endcsname{\def\PYG@tc##1{\textcolor[rgb]{0.10,0.09,0.49}{##1}}}
\expandafter\def\csname PYG@tok@vi\endcsname{\def\PYG@tc##1{\textcolor[rgb]{0.10,0.09,0.49}{##1}}}
\expandafter\def\csname PYG@tok@vm\endcsname{\def\PYG@tc##1{\textcolor[rgb]{0.10,0.09,0.49}{##1}}}
\expandafter\def\csname PYG@tok@sa\endcsname{\def\PYG@tc##1{\textcolor[rgb]{0.73,0.13,0.13}{##1}}}
\expandafter\def\csname PYG@tok@sb\endcsname{\def\PYG@tc##1{\textcolor[rgb]{0.73,0.13,0.13}{##1}}}
\expandafter\def\csname PYG@tok@sc\endcsname{\def\PYG@tc##1{\textcolor[rgb]{0.73,0.13,0.13}{##1}}}
\expandafter\def\csname PYG@tok@dl\endcsname{\def\PYG@tc##1{\textcolor[rgb]{0.73,0.13,0.13}{##1}}}
\expandafter\def\csname PYG@tok@s2\endcsname{\def\PYG@tc##1{\textcolor[rgb]{0.73,0.13,0.13}{##1}}}
\expandafter\def\csname PYG@tok@sh\endcsname{\def\PYG@tc##1{\textcolor[rgb]{0.73,0.13,0.13}{##1}}}
\expandafter\def\csname PYG@tok@s1\endcsname{\def\PYG@tc##1{\textcolor[rgb]{0.73,0.13,0.13}{##1}}}
\expandafter\def\csname PYG@tok@mb\endcsname{\def\PYG@tc##1{\textcolor[rgb]{0.40,0.40,0.40}{##1}}}
\expandafter\def\csname PYG@tok@mf\endcsname{\def\PYG@tc##1{\textcolor[rgb]{0.40,0.40,0.40}{##1}}}
\expandafter\def\csname PYG@tok@mh\endcsname{\def\PYG@tc##1{\textcolor[rgb]{0.40,0.40,0.40}{##1}}}
\expandafter\def\csname PYG@tok@mi\endcsname{\def\PYG@tc##1{\textcolor[rgb]{0.40,0.40,0.40}{##1}}}
\expandafter\def\csname PYG@tok@il\endcsname{\def\PYG@tc##1{\textcolor[rgb]{0.40,0.40,0.40}{##1}}}
\expandafter\def\csname PYG@tok@mo\endcsname{\def\PYG@tc##1{\textcolor[rgb]{0.40,0.40,0.40}{##1}}}
\expandafter\def\csname PYG@tok@ch\endcsname{\let\PYG@it=\textit\def\PYG@tc##1{\textcolor[rgb]{0.25,0.50,0.50}{##1}}}
\expandafter\def\csname PYG@tok@cm\endcsname{\let\PYG@it=\textit\def\PYG@tc##1{\textcolor[rgb]{0.25,0.50,0.50}{##1}}}
\expandafter\def\csname PYG@tok@cpf\endcsname{\let\PYG@it=\textit\def\PYG@tc##1{\textcolor[rgb]{0.25,0.50,0.50}{##1}}}
\expandafter\def\csname PYG@tok@c1\endcsname{\let\PYG@it=\textit\def\PYG@tc##1{\textcolor[rgb]{0.25,0.50,0.50}{##1}}}
\expandafter\def\csname PYG@tok@cs\endcsname{\let\PYG@it=\textit\def\PYG@tc##1{\textcolor[rgb]{0.25,0.50,0.50}{##1}}}

\def\PYGZbs{\char`\\}
\def\PYGZus{\char`\_}

\def\PYGZsh{\char`\#}

\def\PYGZsq{\char`\'}

\makeatother

\usepackage{arxiv}

\usepackage[utf8]{inputenc} 
\usepackage[T1]{fontenc}    
\usepackage{hyperref}       
\usepackage{url}            
\usepackage{booktabs}       
\usepackage{amsfonts}       
\usepackage{nicefrac}       
\usepackage{microtype}      
\usepackage{lipsum}
\usepackage[draft]{minted}
\usepackage{verbatim} 
\usepackage{graphicx}

\title{Music Embedding: A Tool for Incorporating Music Theory into Computational Music Applications}

\author{
  SeyyedPooya HekmatiAthar \\
  Department of Computer Science\\
  North Carolina A\&T State University\\
  Greensboro, NC 27411 \\
  \texttt{shekmatiathar@aggies.ncat.edu} \\
   \And
  Mohd Anwar \\
  Department of Computer Science\\
  North Carolina A\&T State University\\
  Greensboro, NC 27411 \\
  \texttt{manwar@ncat.edu} \\
}

\begin{document}
\maketitle

\begin{abstract}
Advancements in the digital technologies have enabled researchers to develop a variety of Computational Music applications. Such applications are required to capture, process, and generate data related to music. Therefore, it is important to digitally represent music in a music theoretic and concise manner. Existing approaches for representing music are ineffective in terms of utilizing music theory. In this paper, we address the disjoint of music theory and computational music by developing an opensource representation tool based on music theory. Through the wide range of use cases, we run an analysis on the classical music pieces to show the usefulness of the developed music embedding.
\end{abstract}


\section{Introduction}\label{sec:introduction}

No one can say for sure when did music become a conscious part of human development, but undoubtedly, humans have been deliberately creating music for tens of thousands of years \cite{galpin2004howoldismusic}. This old art form has come a long a way to generate more than 50  billion U.S. dollars revenue in our modern age \cite{statista_music_revenue}.  

Similar to many other domains, development of electronics and computers in the $20^{th}$ century has started a new era in the history of music. In this era, not only music-related tasks are being automated, but we are also witnessing new emerging tasks which were difficult to do, if not impossible, a few decades ago. For instance, in the past decade genre identification \cite{Doraisamy2008study}, music summarization \cite{mardirossian2006music}, music database querying \cite{eric2003name}, melodic segmentation \cite{pearce2008comparison}, and harmonic analysis \cite{chen2011music} are presented to the literature. Furthermore, algorithmic tools are used to model patterns of psycho-physical responses to music stimuli \cite{juslin2008emotional}, and the interaction between musical concepts and their interpretations in the brain \cite{krumhansl2001cognitive}. Moreover, using AI and Machine Learning, computers are now capable of generating new pieces of music which is never heard before \cite{dong2018musegan,Huang2018improved}.

Computational music algorithms need to accept music as an input, process it, and provide music as the output. The performance of these algorithms can be improved if we represent the music more effectively. An ineffective representation of music data can become the bottleneck of a system, for instance, the existing generative models struggle to generate music for a large orchestra \cite{guan2019gan}. As it will be discussed in Section~\ref{sec:representation}, the existing representations are developed with pure computational concepts and ignoring high-level music theory. Therefore, the need for a representation which supports the application of high-level music theory concepts is significant. In this paper, we address the disjoint of music theory and computational music by developing an open-source representation tool based on music theory. We then use our tool for studying and analyzing the trends in different composers’ works.

The rest of the paper is organized as follows: Common approaches in representing music and their pros and cons are described in Section~\ref{sec:representation}. Section~\ref{sec:interval} explains intervals in music theory, as a baseline for the solution provided in Section~\ref{sec:music_embedding}. Finally, we demonstrate an example use case of the developed solution in Section~\ref{sec:usecase}, and we conclude the paper in Section~\ref{sec:conclusion}.

\section{Music representation}\label{sec:representation}
In this section we discuss the common practices in representing and storing music in digital media. There are two main categories for data representation in music: audio and symbolic.

\subsection{Audio representation}
Music is an audio which satisfies some specific characteristics and criteria. Therefore, it is reasonable to represent and store music using general audio formats such as WAV and MP3. An analogy to this approach is to store text data in a raster image file, e.g., JPEG; the result can be interpreted and read by human, but the text data is lost. While this approach works great for recording, it does not provide any information about the structure of the music. Therefore, it is not suitable for applications which require data regarding the structure of the music.

\subsection{Symbolic representation}\label{subsec:symbolic}

In contrast to audio representation, symbolic representation stores the structure of the music itself rather than the audio signal which is generated by performance. There exist different approaches for representing music symbolically, each of which providing different level of abstraction. The common theme in these approaches is that music can be seen as a sequence of events. For instance, events in a piece for piano describe when and how hard each piano key should be pressed and how long it should be held. The most common approaches in symbolic music representation are described in the following.

\subsubsection{Music notation}
Music notation is the most human-readable and the oldest representation system. In a standard western notation, music is written using five parallel horizontal lines which are called staff (stave). These five lines represent changes in the pitches, where moving to higher lines increases the pitch. A symbol, called clef, is required to define origin of the pitch and assign notes to lines. The symbols are written and read from left to right and from top to bottom. The shape of each note determines the duration of that note. Usually, vertical lines, called bar lines, intercept staff lines at regular intervals to form bars. A symbol, called time signature, consisting of two numbers describes how many beats each bar represents. 

Music notation was developed and improved with the music itself. Although some scorewriter programs, such as MuseScore and Finale, are developed to create and edit files using music notation, this writing system is designed for human; computers cannot run algorithms efficiently using music notation.

\subsubsection{**kern}
**kern is a representation scheme employed in Humdrum toolkit which is  a set of resources for computational music analysis \cite{Huron2002music}. **kern offers a one-to-one mapping to the basic music notation symbols. **kern can be used to encode pitch and duration as well as common score-related information. In other words, **kern is a text version of music notation, with some limitations.

\subsubsection{ABC notation}
Similar to **kern, ABC notation is a text-based encoding scheme for music notation. ABC notation has less vocabulary than **kern and supports fewer music symbols, but it is easier to read and understand by human.

\subsubsection{MusicXML}
As its name suggests, MusicXML is a music interchange language based on Extensible Markup Language (XML) \cite{muiscXML}. MusicXML inherits pros and cons of XML, namely, MusicXML offers openness, extensibility, self-description, content and presentation separation, while it is verbose and redundant, and more difficult for human to read and understand compared to **kern and ABC notation.

\subsubsection{MIDI standard}
Musical Instrument Digital Interface (MIDI) is the most popular standard for storing and transferring music  \cite{midi}. MIDI standard serves as a storage and retrieval protocol as well as a communication protocol that connects electronic musical instruments, computers, and related audio devices for playing, editing, and recording music. MIDI protocol works based on MIDI messages. A MIDI message describes an event, for instance, a \textit{note on} message signals that a particular note on a particular channel has to sound at the given time with the specified velocity. Due to its binary nature, MIDI is not human-readable, but it is effective in storing and communicating symbolic music between digital interfaces.

\subsubsection{Pianoroll}\label{subsec:pianoroll}
Pianoroll was invented in late $19^{th}$ century as a mechanical mean to store music \cite{welte1883Mechanical}. The general theme for mechanical pianorolls was similar to punched cards: the storage medium is moving under a series of pistons, the holes in the medium make the pistons move which in turn makes the music instrument (generally a piano or a church organ) sound.

With the emergence of Digital Audio Workstations (DAW), pianoroll became digitized. A mechanical pianoroll can be described as a 2D array, in which one dimension represents time while the other one represents notes. The element in the $i^{th}$ row and $j^{th}$ column describes the behavior of the $i^{th}$ note at the $j^{th}$ timestep, where zero means silence and any positive number indicates the velocity of the note.

Pianoroll is the most popular representation scheme in digital music. Popular DAWs such as FL Studio and Cubase only provide pianoroll representation. However, computationally speaking, pianorolls are heavily sparse. This makes it difficult and inefficient to run algorithms using pianorolls.

\section{Intervals in music theory}\label{sec:interval}

As discussed earlier, there is a disjoint between music theory and representation approaches. We aim to close this gap by developing a toolkit based on music theory which can be used to run algorithms efficiently. Since intervals are the building blocks of music theory \cite{piston1948harmony}, it is reasonable to build this toolkit based on intervals. In this section we discuss the intervals while the next section is devoted to the technical details of the tookit.

Music quantizes the sound in both time and frequency domains. In the frequency (pitch) domain, the quantum value is called a semitone. Regardless of different systems to assign a frequency to each semitone (i.e., tuning), everything is measured in terms of semitones. Interval is an inclusive distance measure based on semitones; the relation between two notes is analyzed based on their interval. Interval is the building block of music theory. Almost all high-level musical concepts are built and defined based on intervals. An interval is a multi-dimensional measure, consisting of the following information:
\begin{enumerate}
    \item An ordinal number (usually less than $8^{th}$) which depends on the name of the notes, e.g., A to A is first, A to B is second, B to G is sixth.
    \item A mode/type usually from the set, \{diminished, minor, perfect, major, augmented\}, which depends on the note modifiers, e.g., C to Db is min $2^{nd}$, C to D is maj $2^{nd}$, C to D\# is aug $2^{nd}$.
    \item A direction: ascending or descending (if the second note has a higher pitch it is ascending and vice versa), e.g., C to E is maj $3^{rd}$ asc while E to C is maj $3^{rd}$ des.
    \item Compound versus simple interval. Notes go from A to G then cycle back to A. Two different notes with the same name are called octaves (because they form the $8^{th}$ interval). If an interval is larger than an octave, it is compound and has the same functionality/properties of its corresponding simple interval. For instance, intervals $3^{rd}$ and $10^{th}$ are almost equal.
\end{enumerate}

\begin{table}[tpb]
  \caption{Q-Table for converting semitones to intervals and vice versa}
\centering
  \begin{tabular}{ccccc}
  \toprule
  &\multicolumn{2}{c}{\bfseries Music Theory}& \multicolumn{2}{c}{\bfseries Music Embedding}\\
  \cmidrule(r){2-3} \cmidrule(r){4-5}
  \bfseries Semitones & \bfseries Order & \bfseries  Type  & \bfseries Order & \bfseries Type\\ 
  \midrule
    0&First&Perfect&1&0\\ 
     1&Second&minor&2&-1\\ 
     2&Second&Major&2&1\\ 
     3&Third&minor&3&-1\\
     4&Third&Major&3&1\\ 
     5&Fourth&Perfect&4&0\\ 
     6&Fifth&Diminished&5&-2\\ 
     7&Fifth&Perfect&5&0\\ 
     8&Sixth&minor&6&-1\\ 
     9&Sixth&Major&6&1\\ 
     10&Seventh&minor&7&-1\\ 
     11&Seventh&Major&7&1\\ 
  \bottomrule
  \end{tabular}
\label{tab_q_table}
\end{table}

Different intervals might have the same number of semitones, for instance, aug $4^{th}$ and dim $5^{th}$ both have 6 semitones. The decision between alternative naming is made by taking the scale and alternation into account, which needs expert opinion. The most reasonable alternatives for translating semitones to and from intervals are provided in Table~\ref{tab_q_table}. As mentioned before, compound intervals can be broken into a simple interval plus some octave offset. For this reason, Table~\ref{tab_q_table} does not have a row for 12 semitones because it equals octave which can be encoded as perfect $1^{st}$ plus one octave offset.

\section{Music embedding toolkit}\label{sec:music_embedding}

The proposed Music embedding toolkit is developed for Python. Its source code \footnote{Link to music embedding toolkit source code: \url{https://github.com/PooyaHekmati/music_embedding}} and documentation\footnote{Link to music embedding toolkit documentation: \url{https://pooyahekmati.github.io/music_embedding/}} are publicly available. Additionally, this toolkit is published through Python Package Index and can be accessed with:

\begin{Verbatim}[commandchars=\\\{\}]
\PYG{n}{pip} \PYG{n}{install} \PYG{n}{music\PYGZus{}embedding}
\end{Verbatim}

This Python package is developed based on intervals in the music theory. Hence, it has a class called \textit{interval} which can be used to have interval objects. Another class, called \textit{embedder}, is developed based on the \textit{interval} class. The \textit{embedder} class in the \textit{Music Embedding} package provides functionality to convert pianorolls to and from sequences of intervals with the different configurations. It also provides useful utilities such as extracting melody from a pianoroll, applying Run-Length Encoding (RLE) to a sequence of intervals, and bulk processing.

As explained in Section~\ref{sec:interval}, an interval is the relation between two notes with respect to their pitches; yet, this does not explain how those two notes are related to each other in the time domain. While the notes in an interval can be in any point of time, two special cases are famously known as \textit{melodic interval} and \textit{harmonic interval}. A \textit{melodic interval} is when the second note immediately follows the first note. A \textit{harmonic interval} is when the two notes sound together at the same time. A sequence of \textit{melodic intervals} creates \textit{melody} while a sequence of \textit{harmonic intervals} creates \textit{harmony}. In addition to \textit{melodic} and \textit{harmonic intervals}, the relation between each note and the first note of the same bar is also important. This is because the most strong beat in a bar is usually the first beat and the note at the first beat is of significance. The \textit{embedder} class supports all of these three configurations.

\begin{figure}
 \centering
 \includegraphics[width=\textwidth]{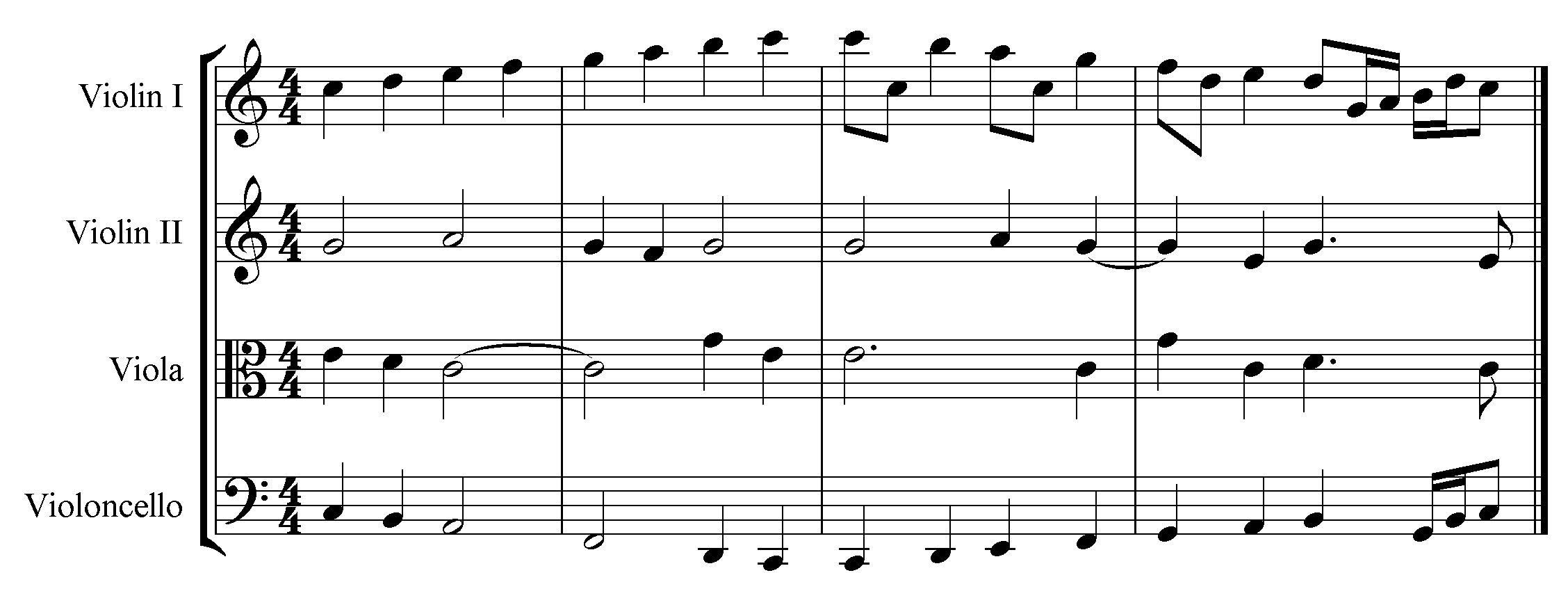}
  \caption{A sample piece for string quartet}
  \label{fig_orchnet_example}
 \end{figure}

\begin{figure}

\centering
\includegraphics[width=\textwidth]{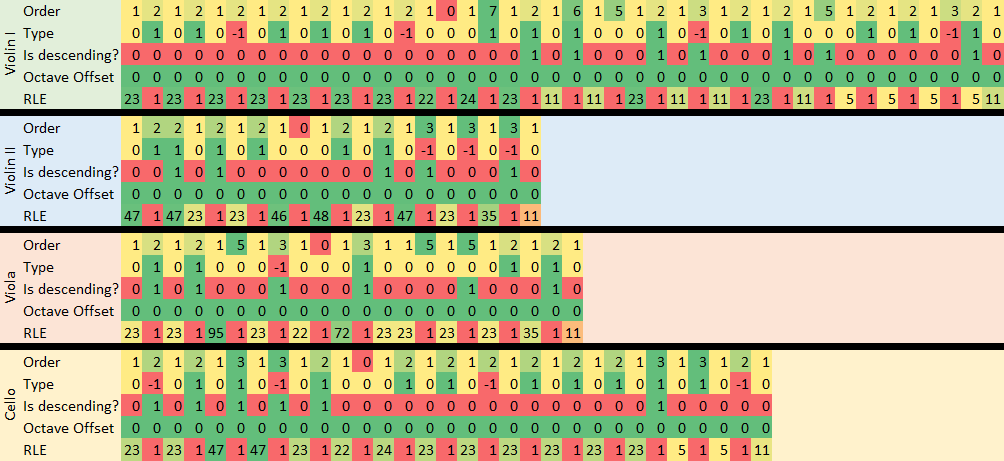}
\caption{The encoded representation of Figure~\ref{fig_orchnet_example} based on intervals}
\label{fig_orchnet_example_encoded}
\end{figure}

Figure~\ref{fig_orchnet_example_encoded} illustrates the melodic embedding of the piece in Figure~\ref{fig_orchnet_example}. The resolution per beat for this example is 24; therefore, each bar has 96 time steps and the entire piece has 384 time steps. In Figure~\ref{fig_orchnet_example_encoded}, consecutive columns which have the same value are summarized into a single column using RLE. This can be interpreted as note duration. The following code snippet is used to generate Figure~\ref{fig_orchnet_example_encoded}.

\begin{Verbatim}[commandchars=\\\{\}]
\PYG{k+kn}{import} \PYG{n+nn}{music\PYGZus{}embedding}
\PYG{k+kn}{import} \PYG{n+nn}{pypianoroll}

\PYG{k}{if} \PYG{n+nv+vm}{\PYGZus{}\PYGZus{}name\PYGZus{}\PYGZus{}} \PYG{o}{==} \PYG{l+s+s1}{\PYGZsq{}\PYGZus{}\PYGZus{}main\PYGZus{}\PYGZus{}\PYGZsq{}}\PYG{p}{:}
    \PYG{n}{midi\PYGZus{}path} \PYG{o}{=} \PYG{l+s+sa}{r}\PYG{l+s+s1}{\PYGZsq{}c:\PYGZbs{}music embedding example.mid\PYGZsq{}}
    \PYG{n}{multi\PYGZus{}track} \PYG{o}{=} \PYG{n}{pypianoroll}\PYG{o}{.}\PYG{n}{read}\PYG{p}{(}\PYG{n}{midi\PYGZus{}path}\PYG{p}{)} \PYG{c+c1}{\PYGZsh{}Open the MIDI file using pypianoroll}
    \PYG{n}{embedder} \PYG{o}{=} \PYG{n}{music\PYGZus{}embedding}\PYG{o}{.}\PYG{n}{embedder}\PYG{o}{.}\PYG{n}{embedder}\PYG{p}{()} \PYG{c+c1}{\PYGZsh{}Create embedder object}
    \PYG{n}{melodic\PYGZus{}intervals} \PYG{o}{=} \PYG{p}{[]}
    \PYG{k}{for} \PYG{n}{i} \PYG{o+ow}{in} \PYG{n+nb}{range}\PYG{p}{(}\PYG{n+nb}{len}\PYG{p}{(}\PYG{n}{multi\PYGZus{}track}\PYG{o}{.}\PYG{n}{tracks}\PYG{p}{)):} \PYG{c+c1}{\PYGZsh{}Get melodic intervals for each track}
        \PYG{n}{melodic\PYGZus{}intervals}\PYG{o}{.}\PYG{n}{append} \PYG{p}{(}\PYG{n}{embedder}\PYG{o}{.}\PYG{n}{get\PYGZus{}melodic\PYGZus{}intervals\PYGZus{}from\PYGZus{}pianoroll} \PYG{p}{(}
            \PYG{n}{pianoroll} \PYG{o}{=} \PYG{n}{multi\PYGZus{}track}\PYG{o}{.}\PYG{n}{tracks}\PYG{p}{[}\PYG{n}{i}\PYG{p}{]}\PYG{o}{.}\PYG{n}{pianoroll}\PYG{p}{))}
    \PYG{n}{melodic\PYGZus{}rle} \PYG{o}{=} \PYG{n}{embedder}\PYG{o}{.}\PYG{n}{get\PYGZus{}RLE\PYGZus{}from\PYGZus{}intervals\PYGZus{}bulk}\PYG{p}{(}\PYG{n}{melodic\PYGZus{}intervals}\PYG{p}{)} \PYG{c+c1}{\PYGZsh{}Get RLE}
\end{Verbatim}

\section{Use cases}\label{sec:usecase}

The music embedding toolkit can be used in numerous applications, including but not limited to statistical and probabilistic analysis of music pieces, developing generative models to have AI-synthesized music, genre classification,
mood recognition, melody extraction, audio-to-score alignment, and score structure analysis. In this section, we explore one simple use case of the music embedding toolkit by analyzing the trend in different composers' works.

We perform the analysis based on a dataset on Kaggle\footnote{Link to the classical dataset on Kaggle: \url{https://www.kaggle.com/blanderbuss/midi-classic-music/}}. This MIDI dataset contains 3,920 pieces from 144 classical composers. After extracting all MIDI files from the zip files in the dataset, we checked the authenticity of the files by randomly selecting files ($n=100$) and comparing them to verified sources. We then performed data cleaning using Pypianoroll\cite{pypianoroll} and pretty\_midi \cite{pretty_midi} packages. The following criteria were considered for excluding a file:
\begin{enumerate}
  \item if either Pypianoroll or pretty\_midi cannot handle the file and throw an error,
  \item if the file has more than one time signature change event,
  \item if the file contains less than 10 bars (the number of bars was calculated as: \\$active\ length/beat\ resolution/time\ signature\\ numerator$),
  \item if the numerator of the first time signature event is 1.
\end{enumerate}

\begin{figure}
  \centering
  \includegraphics[width=0.75\columnwidth]{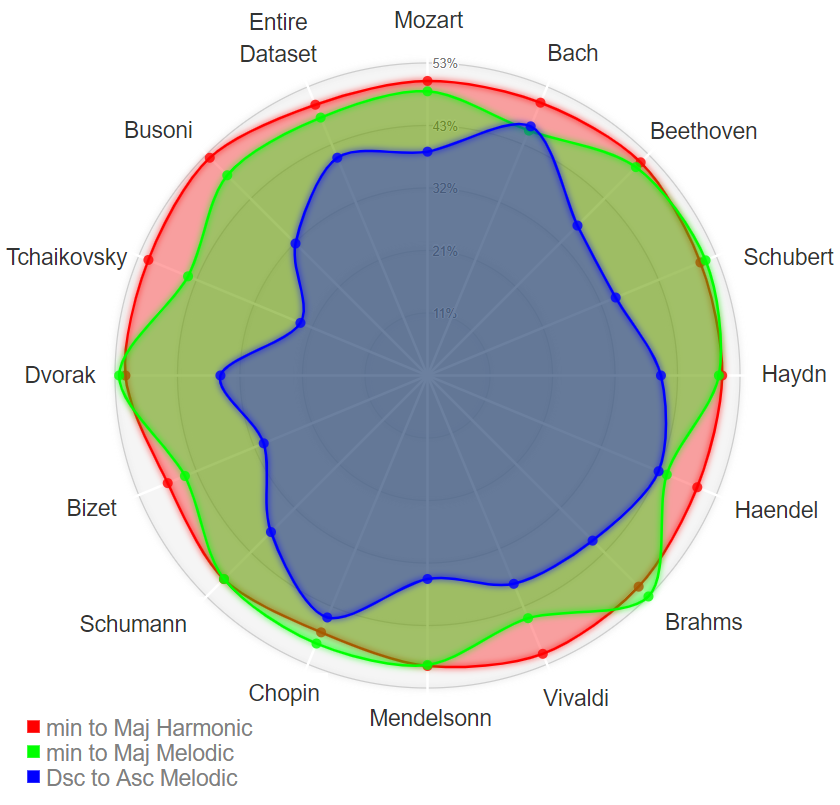}
  \caption{Radar chart for the top 15 composers, minor-to-Major and Descending-to-Ascending ratios}
\label{fig_radar}
\end{figure}

Figure~\ref{fig_radar} shows a radar chart in which 15 composers with the most data in the dataset are compared with each other and the entire dataset. In this chart, the ratio of occurrences of minor and Major intervals are compared both harmonically and melodically. Also, the ratio of descending and ascending melodic intervals is shown. The minor-major ratios are calculated by:
\begin{equation}
r=\frac{f_{min}}{f_{min}+f_{Maj}}
\end{equation}
where $r$ is the ratio and $f_{min}$ and $f_{Maj}$ are the occurrences of minor and Major intervals respectively. Similarly, the ratio of descending and ascending melodic intervals is calculated by:
\begin{equation}
r=\frac{f_{dsc}}{f_{dsc}+f_{asc}}
\end{equation}
where $r$ is the ratio and $f_{dsc}$ and $f_{asc}$ are the occurrences of descending and ascending intervals respectively. 

The minor-Major ratio for harmonic intervals in the entire dataset is 0.5011, which indicates both minor and Major harmonic are equally used in general. Busoni and Tchaikovsky have the highest ratios with 0.5265 and 0.5164 respectively. The lowest ratios belong to Chopin and Bizet with 0.4753 and 0.4807 respectively. This means that the popularity of minor and Major harmonic intervals is almost equal and does not deviate from the general trend from composer to composer.

The minor-Major ratio for melodic intervals in the entire dataset is 0.4773, which is lower than the same value for harmonic intervals. Brahms and Dvorak have the highest melodic ratios with 0.5343 and 0.5273 respectively. The lowest ratios belong to Haendel and Tchaikovsky with 0.4428 and 0.4431 respectively. Although the deviations in the melodic ratios are more than the deviations in the harmonic ratios, still they are within the same range. However, the melodic ratios of only Schubert, Brahms, Chopin, and Dvorak are greater than their harmonic ratios while for the rest of composers and the entire dataset the opposite is true.

The descending to ascending ratio for the entire dataset is 0.4030 which shows that the general trend is to use ascending intervals more often. Since the range of notes is limited, upward and downward movements need to cancel each other out to ensure the notes are in the feasible range. Having a descending to ascending ratio which is far from 0.5 concludes that ascending intervals are generally smaller than descending intervals; therefore, more ascending intervals are needed to compensate. In contrast to Bach who has the highest ratio (0.4611), Tchaikovsky has the lowest ratio (0.2351). This means that on average, out of every four intervals, Tchaikovsky has used only one descending interval.

\begin{figure}[b]
\centering
\includegraphics[width=\textwidth]{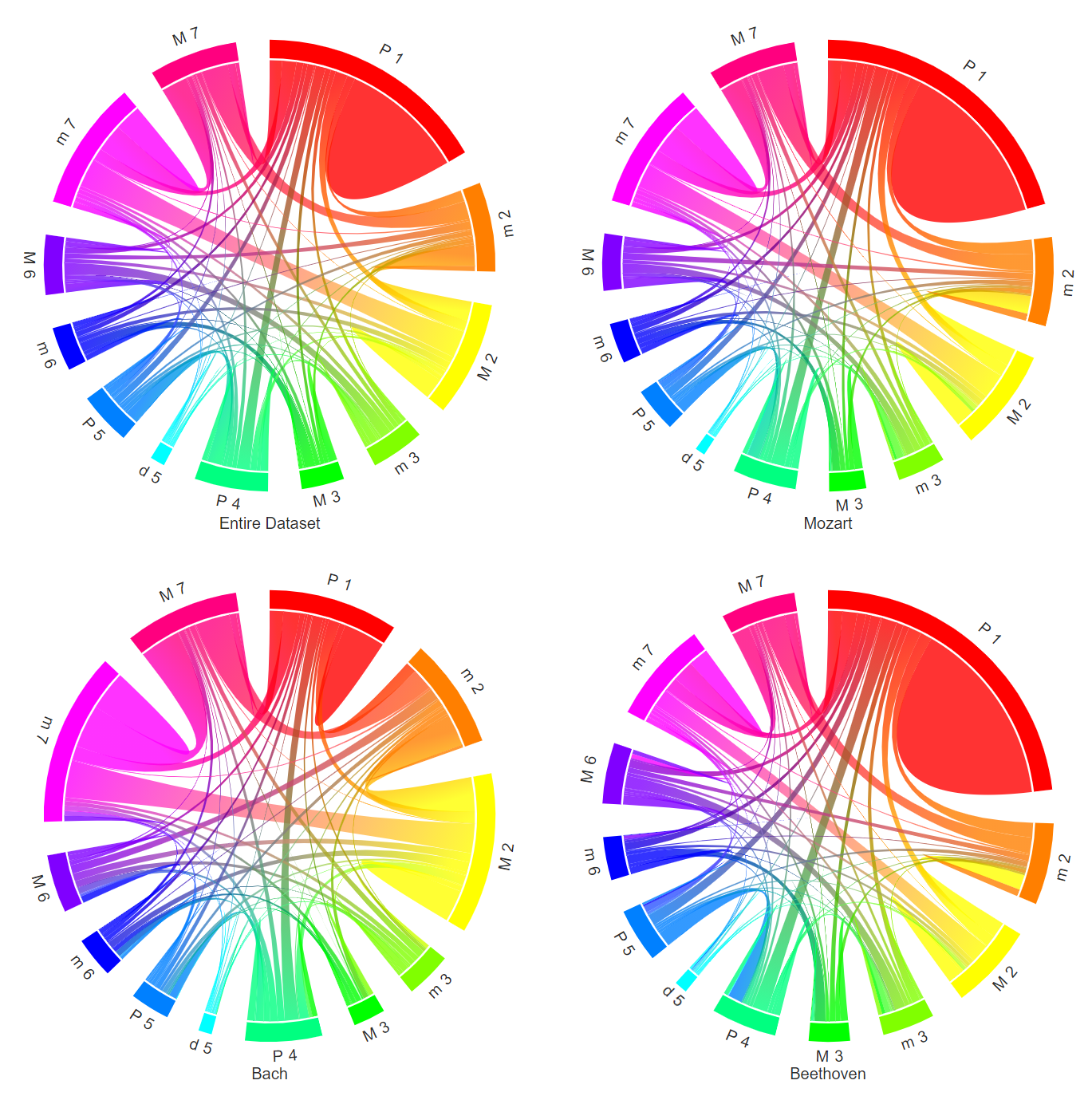}
\caption{Occurrence of pairwise melodic intervals}
\label{fig_melodic_chord}
\end{figure}

\begin{figure}
\centering
\includegraphics[width=\textwidth]{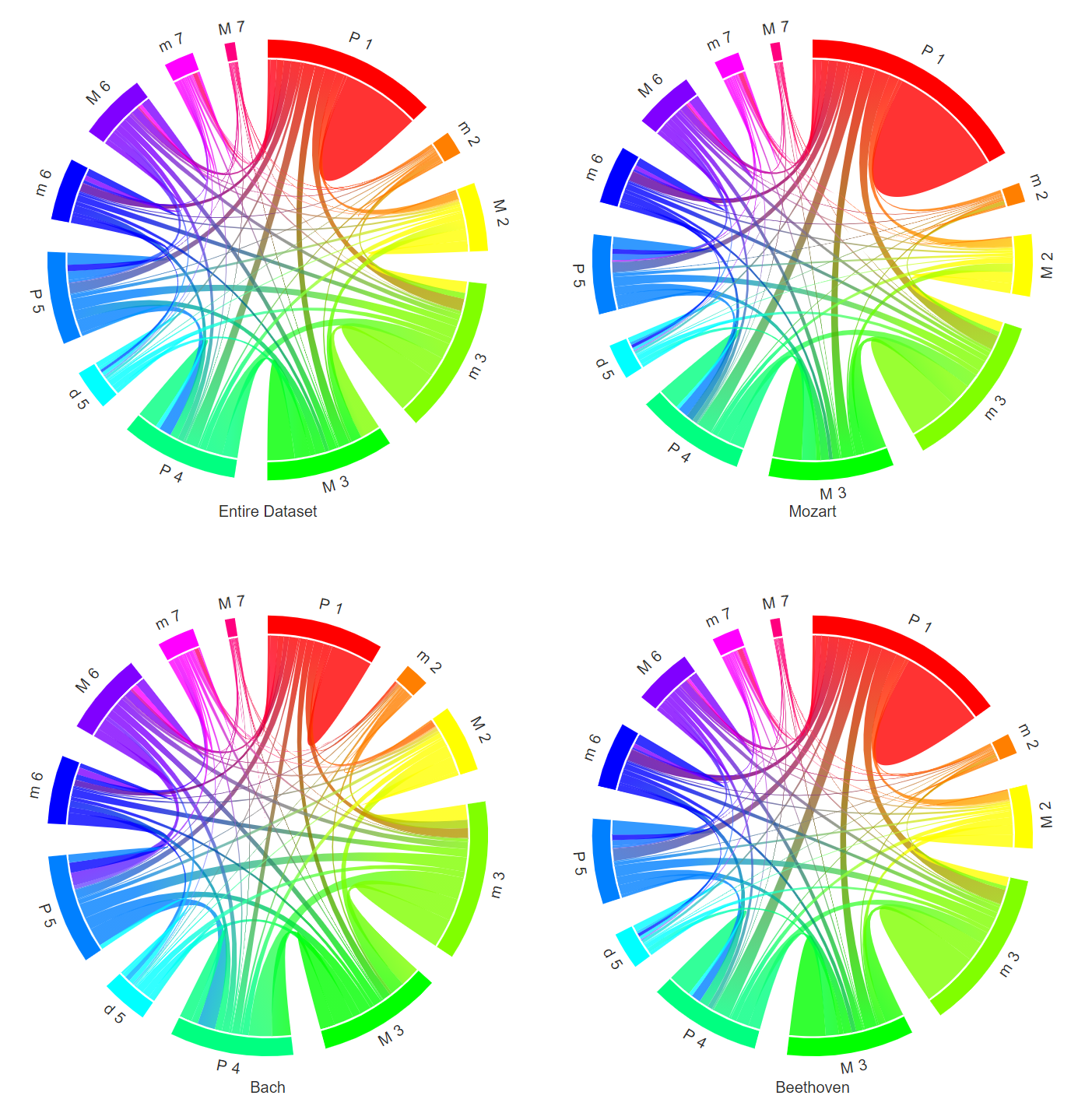}
\caption{Occurrence of pairwise harmonic intervals}
\label{fig_harmonic_chord}
\end{figure}

As explained before, the sequence of intervals makes a piece sound how it sounds. Therefore, analyzing how intervals follow each other is of interest. Figure~\ref{fig_melodic_chord} shows the trends in each pair of melodic intervals happening next to each other. This figure is generated for the entire dataset and Mozart, Bach, and Beethoven as they have the highest number data points in the dataset. Figure~\ref{fig_melodic_chord} shows that unlike Mozart and Beethoven, Bach has used unison and octave intervals less frequently. Also, in all four diagrams, dim $5^{th}$ has the lowest value. Furthermore, the strongest connection is consistently between Maj $2^{nd}$ and min $7^{th}$. A similar figure, but for harmonic intervals is shown in Figure~\ref{fig_harmonic_chord}. Consonant and dissonant intervals can clearly be seen in this figure. Maj $7^{th}$ and min $2^{nd}$ are strongly dissonant in music theory and in Figure~\ref{fig_harmonic_chord} we see how small their proportions are. In contrast, perfect intervals, which are considered to be consonant, have a dominant presence.

\section{Conclusion}\label{sec:conclusion}

There is no doubt that representing data differently can affect efficiency and performance of algorithms. There exist different representation schemes and systems for symbolic music; however, they are developed for storage rather than a communication medium which can be used to let the computers know how humans think when composing or analyzing a musical piece. This shortcoming has slowed down the research and development of computational music domain.

In this paper, we tried to address this gap by presenting the music embedding toolkit which is a Python package based on music theory. This toolkit is developed based on intervals in music theory and offers multiple tools to encode and decode pianorolls to and from sequences of intervals. We showed one potential application of this toolkit by running a numerical analysis on classical pieces of music.

\IfFileExists{\jobname.ent}{
   \theendnotes
}{
}

\bibliographystyle{unsrt}  


\end{document}